\documentclass[]{pasj01}
\draft

\usepackage[utf8]{inputenc}

\Received{}
\Accepted{}
 
\begin{document} 

\title{Tidally Perturbed Oblique Pulsations in the Hierarchical Triple System V1031 Orionis }

\author{Jae Woo \textsc{Lee}\altaffilmark{}%
}
\altaffiltext{}{Korea Astronomy and Space Science Institute, Daejeon 34055, Republic of Korea}
\email{jwlee@kasi.re.kr}

\KeyWords{asteroseismology --- binaries: eclipsing --- stars: fundamental parameters --- stars: individual (V1031 Ori) --- stars: oscillations (including pulsations)}{}

\maketitle

\begin{abstract}
We present TESS photometry of the triple star system V1031 Ori (AB)C, which exhibits short-period oscillations superimposed on 
an eclipsing light curve. The high-quality data were analyzed in detail and combined with the published spectroscopic measurements 
to obtain the fundamental stellar parameters of the program target. The masses and radii of the eclipsing pair (AB) were determined 
to about 0.5 \% and 0.7 \% precision, respectively. We found 23 significant frequencies in two ranges of $<$ 2 day$^{-1}$ and 
10.9$-$12.7 day$^{-1}$ from the eclipse-subtracted residuals. Among them, sixteen in the gravity ($g$)-mode region may be aliases 
and artefacts due to imperfect removal of the systematic trends and the binary effects from the TESS data. Seven frequencies in 
the pressure ($p$)-mode region were separated by the orbital frequency and split by the average offset of 0.042 $\pm$ 0.003 day$^{-1}$ 
from their adjacent harmonics. Further, the pulsation amplitudes are clearly modulated with the binary star orbit. 
The results indicate that the $p$-mode frequencies are tidally perturbed pulsations from the secondary component and the pulsation 
axis could be aligned with the tidal axis.  
\end{abstract}

\section{Introduction}

Detached eclipsing binaries (EBs) can be used to measure the dynamical masses and radii of each component with better than 1 \% 
precision for the best-observed systems (Torres et al. 2010; Maxted et al. 2020), and pulsating stars enable a direct exploration 
of stellar interiors using asteroseismic studies (Aerts et al. 2010; Aerts 2021). Because of the reciprocal information, 
the pulsating EBs are key to our understanding of stellar physics (Murphy 2018; Valle et al. 2018; Tkachenko et al. 2020). 
The tidal interaction between the binary components can cause stellar pulsations that are tidally induced or perturbed. Tidally 
induced modes are pulsations at precise multiples of the orbital frequency ($f_{\rm orb}$). Usually, they are known to oscillate 
in eccentric binary systems when orbital harmonics are equal or close to their eigenmode frequencies (Welsh et al. 2011; 
Hambleton et al. 2013; Guo et al. 2017). Tidally perturbed modes are pulsation modes whose eigen-frequencies are shifted from 
their nominal value in a non-tidally distorted star due to the tidal forces exerted by its companion (Polfliet \& Smeyers 1990; 
Bowman et al. 2019; Southworth et al. 2020). 

This is our third paper about finding and identifying the pulsation features in double-lined EBs from the TESS photometry 
(Lee \& Hong 2021; Lee et al. 2011). We focus on V1031 Ori (TIC 66563761; HD 38735; HIP 27341; Gaia EDR3 3011007889677269888), 
which is a hierarchical triple star system (AB)C composed of a circular eclipsing pair (AB) and an outer companion (C).  
The program target was recognized as an irregular variable by Strohmeier \& Knigge (1962). After that, Andersen \& Nordstr\"om (1977) 
detected the line profiles of three components with approximately equal strength from their spectra of V1031 Ori. Olsen (1977) classified 
the variable as a detached EB with a period of about 3.4057 days from his photometry. Andersen et al. (1990) obtained new 26 spectra 
and complete $uvby$ light curves. Using the spectroscopic data, they measured the radial velocities (RVs) of all three stars, and 
the RV semi-amplitudes ($K_1$, $K_2$) and rotational velocities ($v_1\sin$$i$, $v_2\sin$$i$) of the eclipsing pair. The astrophysical 
parameters for the system were determined by combining these measurements. The result showed that the eclipsing components in 
a circular orbit were spinning much slower than their synchronous rotations. Here, we present the discovery of tidally influenced 
pulsations in the TESS data of V1031 Ori.

\section{Observations and Eclipse Timings}

High-precision photometry of V1031 Ori was carried out in a 2-min cadence mode during Sector 6 of the TESS mission (Ricker et al. 2015). 
The nearly-continuous observations were secured with camera 2 from December 15 2018 to January 6 2019 (BJD 2,458,468.27 $-$ 2,458,490.05). 
We obtained the simple aperture photometry (SAP) measurements called \texttt{SAP$_-$FLUX} from MAST\footnote{https://mast.stsci.edu/portal/Mashup/Clients/Mast/Portal.html} 
and removed four outstanding outliers by visual inspection. To detrend and normalize the raw SAP data, two outside-eclipse light curves 
divided in an intermediate data gap were separately fitted to a second-order polynomial. We converted these fluxes into magnitudes by 
requiring a TESS magnitude of $T\rm_P$ = $+$5.956 (Stassun et al. 2019) at maximum light. The resulting TESS data are depicted in 
Figure 1 as magnitude versus BJD. A total of 14,871 observed points were used for a detailed analysis of this study.

It is possible to get precise minimum epochs from the time-series data. Based on the Kwee \& van Woerden (1956) method, we measured 
12 mid-eclipse timings and their errors presented in columns (1)$-$(2) of Table 1. Then, we applied them to a linear least-squares fit 
to obtain new eclipse ephemeris of V1031 Ori suitable for the TESS observations, as follows: 
\begin{equation}
 \mbox{Min I} = \mbox{BJD}~ (2,458,470.237845\pm0.000017) + (3.4055561\pm0.0000065)E, 
\end{equation}
where the two uncertainties are the 1$\sigma$ values for the ephemeris epoch ($T_0$) and the binary period ($P_{\rm orb}$), and $E$ 
denotes the number of orbit cycles elapsed from this $T_0$. 

In Figure 1, we can see ellipsoidal variations and short-term oscillations, in addition to eclipses. Usually, the former is produced 
by stars that are non-spherical, and the latter comes from pulsating variables. As in the case of IM Per (Lee et al. 2021), 
such oscillations can exert an influence on the timing measurements of V1031 Ori. To compare with the previously-measured timings 
from the observed TESS data, we obtained minimum epochs anew from the pulsation-subtracted data, removing all frequencies detected in 
Section 4, and they are presented in columns (4)$-$(5) of Table 1. As illustrated in the last column of Table 1, the timing measurements 
from both datasets are consistent with each other within 10 sec, and there is no difference between the two minimum types (Min I and Min II).

\section{Binary Modeling}

In order to model the TESS light curve of V1031 Ori, the Wilson-Devinney (W-D) code (Wilson \& Devinney 1971; van Hamme \& Wilson 2007) 
was applied in a scheme identical to that for binary stars CW Cep (Lee \& Hong 2021) and IM Per (Lee et al. 2021). In this modeling, 
we used the mass ratio of $q$ = $K_1/K_2$ = 1.0816$\pm$0.0041 and the surface temperature of $T_{\rm eff,2}$ = 7,850 $\pm$ 500 K, 
obtained from the high-dispersion spectra and multiband light curves of Andersen et al. (1990). Here, subscript 1 denotes the less massive 
but hotter primary component, while the reverse refers to subscript 2. The bolometric albedo and the gravity darkening were fixed to 
the standard values of $A_{1,2}$ = 1.0 and $g_{1,2}$ = 1.0 expected for stars with a radiative envelope. The logarithmic limb-darkening 
law was adopted for bolometric ($X_{\rm bol}$, $Y_{\rm bol}$) and monochromatic ($x_{T_{\rm P}}$, $y_{T_{\rm P}}$) coefficients using 
the tables of van Hamme (1993). Because there was no significant evidence of an orbital eccentricity in the precise TESS data, 
our target star was assumed to be in a circular orbit.

In short-period circular binaries, the rotation rates of the component stars are usually thought to be synchronized to their orbital 
motion due to tidal interaction (Mazeh 2008). For V1031 Ori, Andersen et al. (1990) measured the rotational velocities of the primary 
and secondary stars to be $v_1\sin$$i$ = 22 $\pm$ 2 km s$^{-1}$ and $v_2\sin$$i$ = 43 $\pm$ 4 km s$^{-1}$ from their spectra. 
These values are considerably slower than synchronous rotations of $v_{\rm 1,sync}$ = 44 $\pm1$ km s$^{-1}$ and $v_{\rm 2,sync}$ 
= 64 $\pm1$ km s$^{-1}$ reported by them. Thus, we set the rotation parameters of both components in the W-D code to be their values 
($F_1$ = 0.50, $F_2$ = 0.67) at the beginning, and ours ($F_1$ = 0.52, $F_2$ = 0.66) calculated in this paper at the end. 

The binary modeling for V1031 Ori was conducted by simultaneously solving all TESS data with the W-D differential corrections (DC). 
The DC routine was repeatedly run until the corrections of the free parameters became lower than their standard deviations. 
Our best-fit parameters are presented in Table 2, where $i$ and $\Omega$ are the orbital inclination and the dimensionless surface 
potential, respectively. The parameter errors were obtained following the procedure of Southworth et al. (2020), 
as in our previous works (Lee \& Hong 2021; Lee et al. 2021). The synthetic curve from our result is displayed as a red curve in 
the upper panel of Figure 2 and the corresponding residuals are plotted in the lower panel. As illustrated in the figure, 
the model curve agrees well with the high-quality TESS data and the oscillation feature in the residuals is more clearly seen. 
As a further test, we treated the eccentricity of the eclipsing pair as a free parameter and found it to remain zero.

\section{Pulsational Characteristics}

The residual light curve of V1031 Ori distributed in BJD is plotted in Figure 3. The residuals were formed by subtracting our EB model 
from the observed time-series data. Multiperiodicity is clearly visible in the lower panel of this figure, where the vertical solid and 
dotted lines indicate the primary and secondary minimum epochs, respectively. Moreover, this feature seems to rely on 
the binary orbital phase. Considering the physical properties of the eclipsing pair presented in the next Section, both components lie in 
the $\delta$ Sct instability region on the HR diagram, and the multiperiodic oscillations may come from a pulsating component in the system. 

In order to find the oscillation frequencies of V1031 Ori, we used the PERIOD04 of Lenz \& Breger (2005) in the out-of-eclipse part 
(orbital phases 0.073$-$0.427 and 0.573$-$0.927) of the eclipse-subtracted light residuals. Applying the prewhitening procedure 
(Lee et al. 2014) up to the Nyquist limit of 360 day$^{-1}$, we extracted 23 significant frequencies with a S/N amplitude ratio larger 
than 4.0 for each peak. The results from the multiple frequency analyses are given in Table 3. The uncertainties of the parameters in 
the table were estimated following Kallinger et al. (2008). The Fourier fit obtained from PERIOD04 was overplotted as a red solid curve 
in the lower panel of Figure 3. The amplitude spectra for V1031 Ori are presented in Figure 4. Of the 23 frequencies extracted, sixteen 
are located in the gravity ($g$)-mode region of $<$ 2 day$^{-1}$ and the other seven are in the pressure ($p$)-mode region, from 10 to 
13 day$^{-1}$. 

The Rayleigh frequency of $R_{\rm f}$ = 1/$\Delta T$ ($\Delta T$ is the time interval of data used) is a minimum frequency that can 
be used to resolve pulsation signals. We looked carefully for possible combination frequencies ($nf_i \pm mf_j$) and orbital multiples 
($nf_{\rm orb}$) within the frequency resolution of 1.5$R_{\rm f}$ = 0.068 day$^{-1}$ (Loumos \& Deeming 1978), where $n$ 
and $m$ are integers and $f_{\rm orb}$ is an orbital frequency of 0.293635 day$^{-1}$. The results from this process are discussed in 
the last column of Table 3. Most of the low-frequency signals would not be $g$ modes but could be instrumental artefacts and aliases, 
which are caused by insufficient removal of the systematic trends and the binary effects from the observed data. On the other hand, 
the seven high frequencies correspond to the short periods ($P_{\rm pul}$) of 0.078$-$0.092 days and the pulsation constants ($Q$) of 
0.013$-$0.016 days, which are in the $p$ modes of $\delta$ Sct pulsating stars (Breger 2000; Aerts et al. 2010; Antoci et al. 2019).

\section{Discussion and Conclusions}

The TESS high-quality data of the triple star V1031 Ori exhibits multiple oscillations superimposed on eclipse and ellipsoidal effects. 
Our binary modeling confirmed Andersen et al.'s (1990) study that concluded V1031 Ori is a circular-orbit detached EB. The fill-out factors 
$\Omega_{1,2}$/$\Omega_{\rm in}$ of the primary and secondary stars were about 59 \% and 80 \%, respectively. To determine the physical 
properties of the eclipsing pair, we followed the procedure applied by Lee et al. (2021). New light-curve parameters found in this work were 
combined with the RV semi-amplitudes ($K_1$ = 123.23 $\pm$ 0.32 km s$^{-1}$, $K_2$ = 113.93 $\pm$ 0.31 km s$^{-1}$) of Andersen et al. (1990). 
The fundamental parameters are shown in Table 4, together with those of Andersen et al. (1990). We computed the synchronized rotations of 
$v_{\rm 1,sync}$ = 42.21 $\pm$ 0.30 km s$^{-1}$ and $v_{\rm 2,sync}$ = 65.41 $\pm$ 0.47 km s$^{-1}$. These indicate that both components 
are slowly rotating by a factor of 0.52 and 0.66 in the same order as before. 

Considering $V$ = 6.017 $\pm$ 0.005 mag at 0.25 orbital phase (Max I) (Andersen et al. 1990) and our light ratio ($l_{1,2,3}$) and 
ignoring interstellar extinction ($A_{\rm V}$ = 0), we obtained a distance to our target star V1031 Ori to be 210 $\pm$ 26 pc. 
This is in excellent agreement with 215 $\pm$ 25 pc calculated by Andersen et al. (1990) and 200 $\pm$ 40 pc taken by 
the Hipparcos parallax (5.0$\pm$1.0 mas, ESA 1997), while it is much closer than the geometric and photogeometric distances of 
$588_{+389}^{-198}$ pc and $734_{+290}^{-258}$ pc, respectively, estimated by Bailer-Jones et al. (2021) from the GAIA EDR3 source 
($\pi$ = 2.5 $\pm$ 1.0 mas, $G$ = 6.089 $\pm$ 0.007, $BP-RP$ = 0.221 $\pm$ 0.009; Gaia Collaboration et al. 2020). 
 
We applied a multifrequency analysis to the residual light curve after subtracting the binary model. This resulted in the detection 
of 23 oscillation frequencies with S/N $>$ 4.0, of which the high frequencies between 10$-$13 day$^{-1}$ are identified as $p$ modes 
from their periods and pulsation constants. The pulsations in close binaries can be excited or perturbed by tidal effects between 
the component stars (Polfliet \& Smeyers 1990; Welsh et al. 2011; Hambleton et al. 2013; Guo et al. 2017; Bowman et al. 2019; 
Southworth et al. 2020). Because the eclipsing pair of V1031 Ori are in a circular orbit, we can hardly expect tidally excited modes 
in the EB system. The bottom panel of Figure 4 is the amplified spectrum of a section of the high-frequency region after prewhitening 
the two highest signals $f_1$ and $f_2$, marked as blue solid lines. As displayed in this panel, the pulsation signals are separated 
by the orbital frequency and split by a constant interval from their adjacent harmonics. 
We subtracted the orbital harmonics ($nf_{\rm orb}$) from the seven high frequencies ($f_{1,2,4,7,10,12,18}$). The frequency differences 
are presented in the seventh column of Table 3, from which we obtained the average frequency offset of 0.042 $\pm$ 0.003 day$^{-1}$. 
These demonstrate that the multiperiodic oscillations in V1031 Ori are tidally perturbed modes, similar to those exhibited in 
the TESS data of the intermediate-mass semi-detached Algol U Gru (Bowman et al. 2019) and the high-mass detached binaries V453 Cyg 
(Southworth et al. 2020) and VV Ori (Southworth et al. 2021). 

In Figures 2 and 3, the light curve residuals of V1031 Ori display amplitude modulations with the orbital phase. Specifically, 
the oscillation amplitudes are the largest in the primary eclipse and definitely diminish in the secondary eclipse and the quadratures 
of Max I and Max II. These indicate that the amplitudes of the observed pulsation modes vary with the binary orbit, i.e., 
the visible surface area. From our binary model and the TESS light curve, at the primary minima the larger and more massive secondary 
completely occults the primary component, while at the secondary minima the hotter but smaller primary transits the secondary star. 
Thus, the amplitude reduction in the secondary eclipse may result from partial eclipses of the pulsating secondary star by its companion. 
Also, the amplitude reduction in both quadratures could be explained by considering that the pulsation axis is aligned with the line of 
apsides (i.e., the tidal axis) connecting the binary components, and hence precesses with the binary star orbit. As in the cases of 
HD 74423 (Handler et al. 2020) and CO Cam (Kurtz et al. 2020), it is possible that the tidal distortion on the oscillating secondary 
tilts its pulsation axis and causes the observed amplitude modulations. Then, the secondary component of the program target could be 
an oblique pulsator, which is useful for studying the influence of tidal distortion on stellar pulsation. The discovery of the tidally 
perturbed oblique pulsations makes the hierarchical triple system V1031 Ori an ideal specimen for tidal asteroseismology. 
High-resolution spectroscopy will greatly help measure the precise $T_{\rm eff}$ and $v\sin$$i$ of each component and detect the changes 
in line profiles and RVs of a pulsating variable and a circumbinary object.

\begin{ack}
This paper includes data collected by the TESS mission, which were obtained from MAST. Funding for the TESS mission is provided by 
the NASA Explorer Program. The authors wish to thank the TESS team for its support of this work. This research has made use of 
the Simbad database maintained at CDS, Strasbourg, France, and was supported by the KASI grant 2021-1-830-08. 
\end{ack}

\clearpage
\begin{figure}
\includegraphics[scale=0.9]{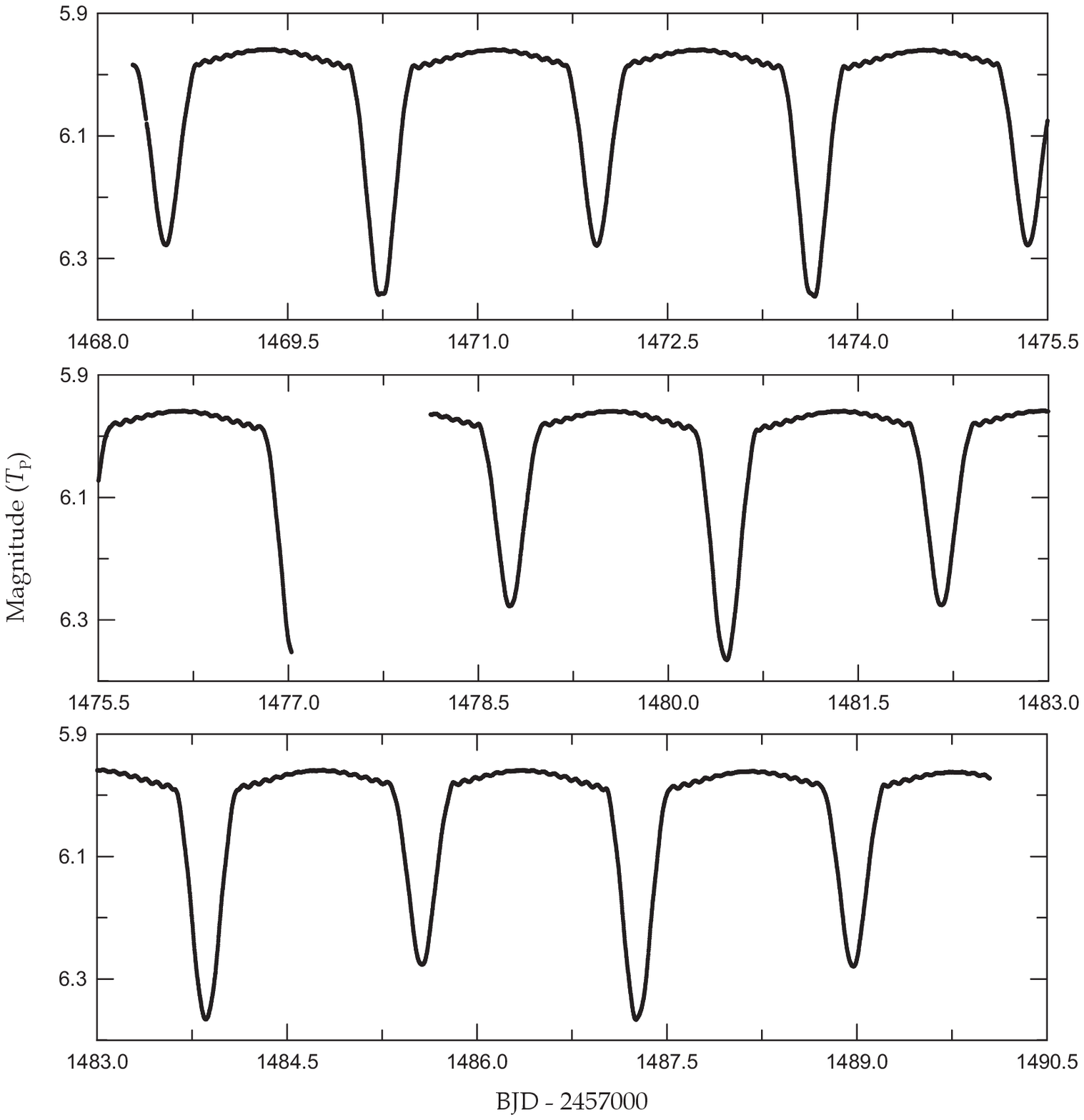}
\caption{TESS time-series data of V1031 Ori observed during Sector 6. Short-term oscillations are clearly shown in the out-of-eclipse parts. }
\label{Fig1}
\end{figure}

\begin{figure}
\includegraphics[]{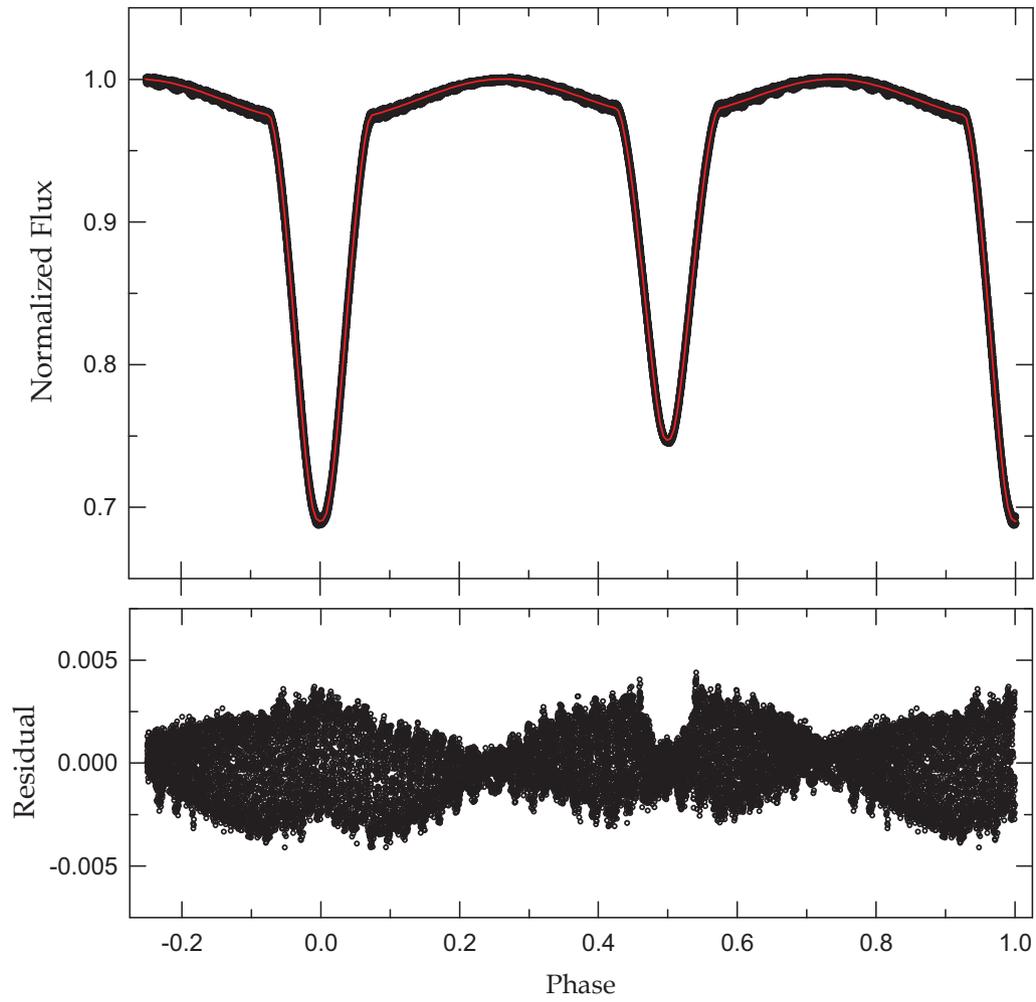}
\caption{Phase-folded light curve of V1031 Ori with the fitted model. In the upper panel, the red line represents the synthetic curve 
obtained from our binary model in Table 2. The corresponding residuals are plotted in the lower panel. }
\label{Fig2}
\end{figure}

\begin{figure}
\includegraphics[]{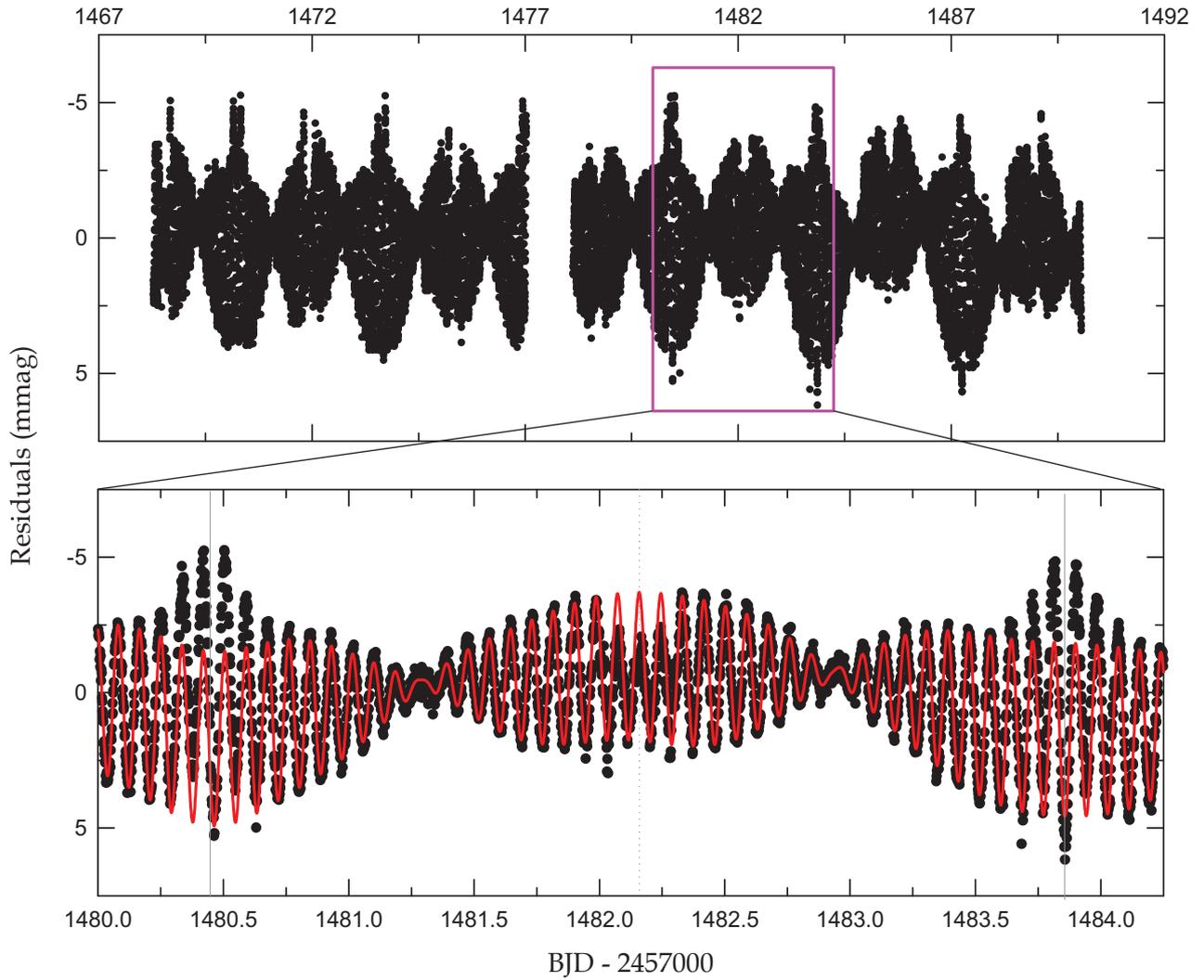}
\caption{Light curve residuals distributed in BJD after removing the binary model. The lower panel is a short section of the residuals 
marked using the inset box of the upper panel. The vertical solid and dotted lines indicate the primary and secondary minimum epochs, 
respectively, and the synthetic curve is computed from the 23-frequency fit to the outside-eclipse data.}
\label{Fig3}
\end{figure}

\begin{figure}
\includegraphics[]{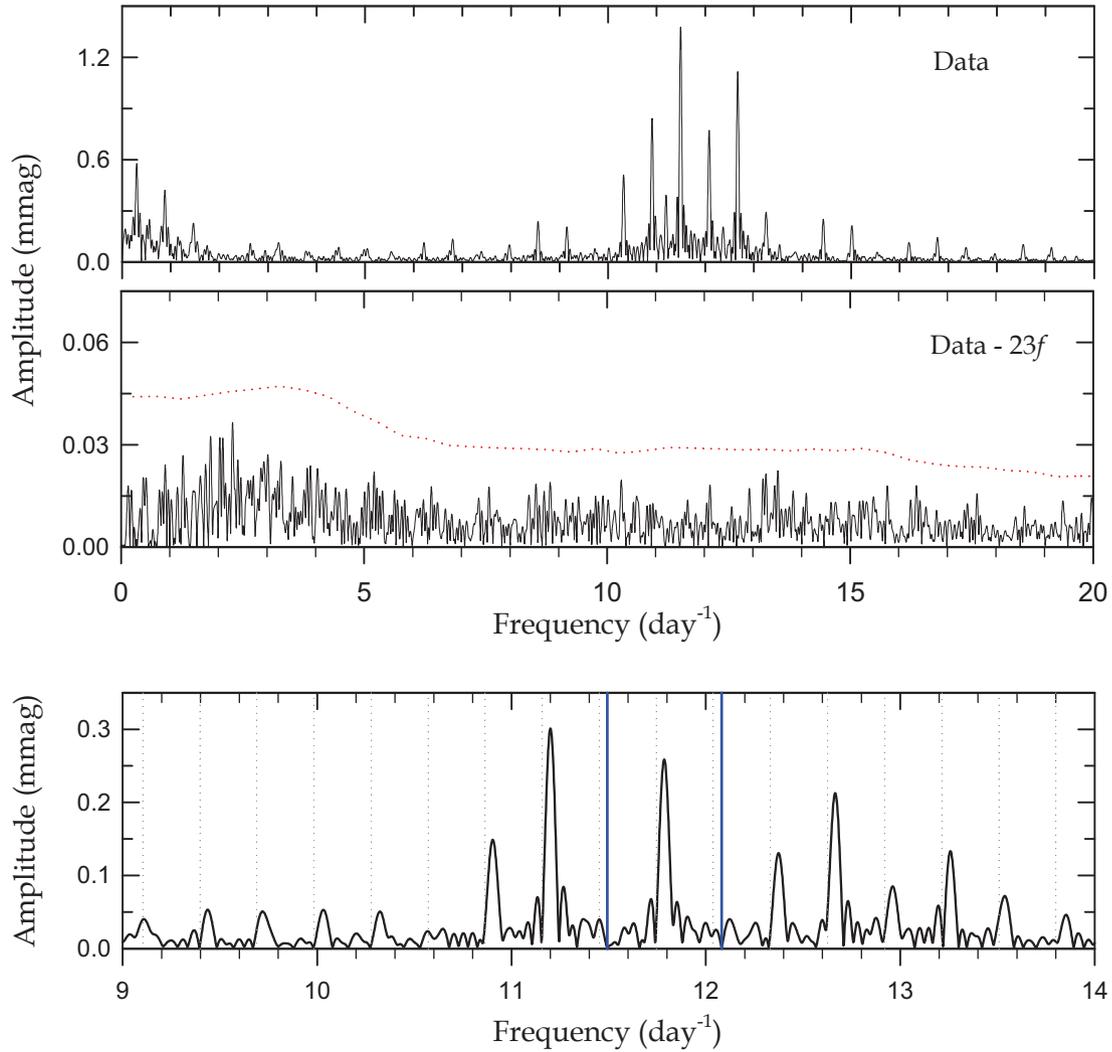}
\caption{Periodogram from the PERIOD04 program for the entire outside-eclipse data. The amplitude spectra before and after prewhitening 
all 23 frequencies are shown in the top and middle panels, respectively. The dotted line in the middle panel corresponds to four times 
the noise spectrum. The bottom panel is the amplified spectrum of a section of the high-frequency region after prewhitening 
the two highest signals $f_1$ and $f_2$, which are plotted as blue solid lines. The vertical dotted gray lines denote the multiples of 
the orbital frequency. }
\label{Fig4}
\end{figure}

\clearpage 
\begin{table}
\tbl{Eclipse Timings of V1031 Ori Measured from Both Datasets. }{%
\begin{tabular}{lccccc}
\hline
\multicolumn{2}{c}{Observed Data}   &                & \multicolumn{2}{c}{Pulsation-subtracted Data}  & Difference$\rm ^a$    \\ [1.0mm] \cline{1-2} \cline{4-5} 
BJD              & Error            & Min            & BJD              & Error                       &                       \\
\hline                                         
2,458,468.53511  & $\pm$0.00006     & II             & 2,458,468.53504  & $\pm$0.00004                & $+$0.00007            \\
2,458,470.23786  & $\pm$0.00002     & I              & 2,458,470.23777  & $\pm$0.00001                & $+$0.00009            \\
2,458,471.94056  & $\pm$0.00004     & II             & 2,458,471.94065  & $\pm$0.00003                & $-$0.00009            \\
2,458,473.64331  & $\pm$0.00011     & I              & 2,458,473.64342  & $\pm$0.00002                & $-$0.00011            \\
2,458,475.34603  & $\pm$0.00011     & II             & 2,458,475.34607  & $\pm$0.00006                & $-$0.00004            \\
2,458,478.75168  & $\pm$0.00011     & II             & 2,458,478.75175  & $\pm$0.00005                & $-$0.00007            \\
2,458,480.45463  & $\pm$0.00010     & I              & 2,458,480.45452  & $\pm$0.00002                & $+$0.00011            \\
2,458,482.15729  & $\pm$0.00002     & II             & 2,458,482.15726  & $\pm$0.00001                & $+$0.00003            \\
2,458,483.86008  & $\pm$0.00004     & I              & 2,458,483.86009  & $\pm$0.00001                & $-$0.00001            \\
2,458,485.56282  & $\pm$0.00008     & II             & 2,458,485.56291  & $\pm$0.00005                & $-$0.00009            \\
2,458,487.26566  & $\pm$0.00015     & I              & 2,458,487.26568  & $\pm$0.00003                & $-$0.00002            \\
2,458,488.96849  & $\pm$0.00012     & II             & 2,458,488.96847  & $\pm$0.00006                & $+$0.00002            \\
\hline
\end{tabular}}\label{tab:1}
\begin{tabnote}
\footnotemark[a]Differences between columns (1) and (4). \\
\end{tabnote}
\end{table}

\begin{table}
\tbl{Binary Parameters of V1031 Ori. }{%
\begin{tabular}{lcc}
\hline
Parameter                                & Primary                  & Secondary                  \\                                                                                         
\hline                                                                                      
$T_0$ (BJD)                              & \multicolumn{2}{c}{2,458,470.237771$\pm$0.000057}     \\
$P_{\rm orb}$ (day)                      & \multicolumn{2}{c}{3.405592$\pm$0.000018}             \\
$q$                                      & \multicolumn{2}{c}{1.0816$\pm$0.0041}                 \\
$i$ (deg)                                & \multicolumn{2}{c}{84.647$\pm$0.075}                  \\
$T_{\rm eff}$ (K)                        & 8797$\pm$500             & 7850$\pm$500               \\
$\Omega$                                 & 6.728$\pm$32             & 4.921$\pm$21               \\
$\Omega_{\rm in}$$\rm ^a$                & \multicolumn{2}{c}{3.881}                             \\
$A$                                      & 1.0                      & 1.0                        \\
$g$                                      & 1.0                      & 1.0                        \\
$F$                                      & 0.52                     & 0.66                       \\
$X_{\rm bol}$, $Y_{\rm bol}$             & 0.654, 0.119             & 0.666, 0.162               \\
$x_{T_{\rm P}}$, $y_{T_{\rm P}}$         & 0.510$\pm$0.010, 0.209   & 0.557$\pm$0.014, 0.215     \\
$l$/($l_{1}$+$l_{2}$+$l_{3}$)            & 0.2813$\pm$0.0011        & 0.5284                     \\
$l_{3}$$\rm ^b$                          & \multicolumn{2}{c}{0.1903$\pm$0.0028}                 \\
$r$ (pole)                               & 0.1766$\pm$0.0011        & 0.2707$\pm$0.0018          \\
$r$ (point)                              & 0.1787$\pm$0.0012        & 0.2837$\pm$0.0021          \\
$r$ (side)                               & 0.1768$\pm$0.0011        & 0.2729$\pm$0.0018          \\
$r$ (back)                               & 0.1783$\pm$0.0012        & 0.2799$\pm$0.0020          \\
$r$ (volume)$\rm ^c$                     & 0.1773$\pm$0.0012        & 0.2748$\pm$0.0019          \\
\hline
\end{tabular}}\label{tab:2}
\begin{tabnote}
\footnotemark[a]Potential for the inner critical surface. \\
\footnotemark[b]Value at 0.25 orbital phase. \\
\footnotemark[c]Mean volume radius. \\
\end{tabnote}
\end{table}

\begin{table}
\tbl{Results of the multiple frequency analysis for V1031 Ori$\rm ^a$. }{%
\begin{tabular}{lrccrccc}
\hline
             & Frequency              & Amplitude           & Phase           & S/N            & $n$            & $f-nf_{\rm orb}$$\rm ^b$ & Remark                    \\ [-1.5ex]
             & (day$^{-1}$)           & (mmag)              & (rad)           &                &                & (day$^{-1}$)             &                           \\
\hline                                                                                                                            
$f_{1}$      & 11.49342$\pm$0.00006   & 1.807$\pm$0.012     & 4.60$\pm$0.02   & 250.15         & 39             & 0.0417                   &                           \\  
$f_{2}$      & 12.08130$\pm$0.00008   & 1.350$\pm$0.012     & 3.06$\pm$0.03   & 189.62         & 41             & 0.0423                   &                           \\  
$f_{3}$      &  0.29623$\pm$0.00019   & 0.887$\pm$0.019     & 2.84$\pm$0.06   &  80.77         &                &                          & $f_{\rm orb}$             \\  
$f_{4}$      & 11.19949$\pm$0.00065   & 0.177$\pm$0.013     & 2.68$\pm$0.21   &  24.22         & 38             & 0.0414                   &                           \\  
$f_{5}$      &  0.56491$\pm$0.00050   & 0.345$\pm$0.019     & 0.79$\pm$0.16   &  31.09         &                &                          & $2f_{\rm orb}$            \\  
$f_{6}$      &  0.07578$\pm$0.00069   & 0.251$\pm$0.019     & 4.80$\pm$0.22   &  22.54         &                &                          &                           \\  
$f_{7}$      & 12.66458$\pm$0.00047   & 0.237$\pm$0.012     & 6.17$\pm$0.15   &  33.59         & 43             & 0.0383                   &                           \\  
$f_{8}$      &  0.17912$\pm$0.00108   & 0.162$\pm$0.019     & 5.59$\pm$0.34   &  14.54         &                &                          & $2f_6$                    \\  
$f_{9}$      &  0.33068$\pm$0.00097   & 0.177$\pm$0.019     & 5.02$\pm$0.31   &  16.06         &                &                          & $f_{\rm orb}$             \\  
$f_{10}$     & 11.78507$\pm$0.00044   & 0.259$\pm$0.012     & 1.20$\pm$0.14   &  35.85         & 40             & 0.0397                   &                           \\  
$f_{11}$     &  0.62002$\pm$0.00076   & 0.229$\pm$0.019     & 1.87$\pm$0.24   &  20.69         &                &                          & $2f_{\rm orb}$            \\  
$f_{12}$     & 12.37294$\pm$0.00072   & 0.155$\pm$0.012     & 6.24$\pm$0.23   &  21.67         & 42             & 0.0403                   &                           \\  
$f_{13}$     &  0.04133$\pm$0.00092   & 0.191$\pm$0.019     & 2.92$\pm$0.29   &  17.09         &                &                          & $f_6$                     \\  
$f_{14}$     &  0.24801$\pm$0.00098   & 0.176$\pm$0.019     & 5.17$\pm$0.31   &  15.99         &                &                          & $f_{\rm orb}$             \\  
$f_{15}$     &  0.72566$\pm$0.00165   & 0.105$\pm$0.019     & 4.72$\pm$0.53   &   9.47         &                &                          & $2f_9$                    \\  
$f_{16}$     &  0.93922$\pm$0.00153   & 0.112$\pm$0.019     & 0.80$\pm$0.49   &  10.20         &                &                          & $f_{12}-f_1$              \\  
$f_{17}$     &  0.68432$\pm$0.00219   & 0.078$\pm$0.019     & 4.49$\pm$0.70   &   7.13         &                &                          & $f_{11}$                  \\  
$f_{18}$     & 10.91244$\pm$0.00132   & 0.085$\pm$0.012     & 1.67$\pm$0.42   &  11.89         & 37             & 0.0479                   &                           \\  
$f_{19}$     &  1.69703$\pm$0.00278   & 0.063$\pm$0.019     & 2.70$\pm$0.89   &   5.63         &                &                          & $f_7-f_{18}$              \\  
$f_{20}$     &  1.40309$\pm$0.00247   & 0.070$\pm$0.019     & 4.79$\pm$0.79   &   6.33         &                &                          & $2f_{15}$                 \\  
$f_{21}$     &  1.55695$\pm$0.00348   & 0.050$\pm$0.019     & 5.50$\pm$1.11   &   4.50         &                &                          & $f_{16}+2f_{\rm orb}$     \\  
$f_{22}$     &  1.79348$\pm$0.00348   & 0.050$\pm$0.019     & 2.58$\pm$1.11   &   4.50         &                &                          & $f_{16}+3f_{\rm orb}$     \\  
$f_{23}$     &  0.99893$\pm$0.00300   & 0.057$\pm$0.019     & 5.27$\pm$0.96   &   5.21         &                &                          & $f_{16}$                  \\  
\hline
\end{tabular}}\label{tab:3}
\begin{tabnote}
\footnotemark[a]Frequencies are listed in order of detection. \\
\footnotemark[b]Difference between each frequency ($f$) and its adjacent orbital harmonic ($nf_{\rm orb}$). \\
\end{tabnote}
\end{table}

\begin{table}
\tbl{Absolute Parameters of V1031 Ori. }{%
\begin{tabular}{lcccccc}
\hline
Parameter                    & \multicolumn{2}{c}{Andersen et al. (1990)}  && \multicolumn{2}{c}{This Paper}              \\ [1.0mm] \cline{2-3} \cline{5-6} 
                             & Primary            & Secondary              && Primary            & Secondary              \\                                                                                         
\hline 
$M$ ($M_\odot$)              & 2.287$\pm$0.017    & 2.473$\pm$0.018        && 2.291$\pm$0.011    & 2.478$\pm$0.011        \\
$R$ ($R_\odot$)              & 2.987$\pm$0.064    & 4.321$\pm$0.034        && 2.840$\pm$0.020    & 4.401$\pm$0.032        \\
$\log$ $g$ (cgs)             & 3.850$\pm$0.019    & 3.560$\pm$0.007        && 3.8915$\pm$0.0064  & 3.5449$\pm$0.0065      \\
$v_{\rm sync}$ (km s$^{-1}$) & 44$\pm$1           & 64$\pm$1               && 42.21$\pm$0.30     & 65.41$\pm$0.47         \\
$\rho$ ($\rho_\odot$)        &                    &                        && 0.1002$\pm$0.0022  & 0.0291$\pm$0.0006      \\
$T_{\rm eff}$ (K)            & 8400$\pm$500       & 7850$\pm$500           && 8797$\pm$500       & 7850$\pm$500           \\
$\log$ $L$ ($L_\odot$)       & 1.60$\pm$0.11      & 1.80$\pm$0.11          && 1.64$\pm$0.10      & 1.82$\pm$0.11          \\
$M_{\rm bol}$ (mag)          & 0.70$\pm$0.26      & 0.18$\pm$0.28          && 0.64$\pm$0.25      & 0.18$\pm$0.28          \\
BC (mag)                     & $-$0.04            & 0.00                   && $-$0.03$\pm$0.06   & 0.02$\pm$0.01          \\
$M_{\rm V}$ (mag)            & 0.74$\pm$0.26      & 0.18$\pm$0.28          && 0.67$\pm$0.26      & 0.16$\pm$0.28          \\
Distance (pc)                & \multicolumn{2}{c}{215$\pm$25}              && \multicolumn{2}{c}{210$\pm$26}              \\
\hline
\end{tabular}}\label{tab:4}
\end{table}

\end{document}